\documentclass[fleqn,usenatbib]{mnras}

\usepackage[T1]{fontenc}

\DeclareRobustCommand{\VAN}[3]{#2}
\let\VANthebibliography\thebibliography
\def\thebibliography{\DeclareRobustCommand{\VAN}[3]{##3}\VANthebibliography}

\usepackage{graphicx}	
\usepackage{amsmath}	
\usepackage{amssymb}	

\usepackage[ruled,vlined]{algorithm2e}
\usepackage{xcolor}  
\usepackage{tabularx}
\usepackage{siunitx}

\usepackage{newtxtext,newtxmath}

\title[IQRM: RFI masking for radio transient searches]{IQRM: real-time adaptive RFI masking for radio transient and pulsar searches}

\author[V. Morello et al.]{
V. Morello,$^{1}$\thanks{E-mail: vincent.morello@manchester.ac.uk}
K. M. Rajwade,$^{1}$
and B.W. Stappers$^{1}$
\\
$^{1}$Jodrell Bank Centre for Astrophysics, Department of Physics and Astronomy, The University of Manchester, Manchester M13 9PL \\
}

\date{Accepted XXX. Received YYY; in original form ZZZ}

\pubyear{2021}

\begin{document}
\label{firstpage}
\pagerange{\pageref{firstpage}--\pageref{lastpage}}
\maketitle

\begin{abstract}

In a search for short timescale astrophysical transients in time-domain data, radio-frequency interference (RFI) causes both large quantities of false positive candidates and a significant reduction in sensitivity if not correctly mitigated. Here we propose an algorithm that infers a time-variable frequency channel mask directly from short-duration ($\sim$1~s) data blocks: the method consists of computing a spectral statistic that correlates well with the presence of RFI, and then finding high outliers among the resulting values. For the latter task, we propose an outlier detection algorithm called Inter-Quartile Range Mitigation (IQRM), that is both non-parametric and robust to the presence of a trend in sequential data. The method requires no training and can in principle adapt to any telescope and RFI environment; its efficiency is shown on data from both the MeerKAT and Lovell 76-m radio telescopes. IQRM is fast enough to be used in a streaming search and has been integrated into the MeerTRAP real-time transient search pipeline. Open-source \textsc{python} and \textsc{C++} implementations are also provided.

\end{abstract}



\section{Introduction}


Radio-frequency interference broadly refers to all signals that negatively impact radio astronomical observations. While the vast majority emanate from artificial sources, naturally occurring phenomena such as lightning can regularly affect some observatories as well \citep{Sokolowski2015}. In radio transient and pulsar searches, idealised data consist of the superposition of a weak astrophysical signal with uncorrelated Gaussian noise in all frequency channels. In this case, the theoretically optimal detection method is known both for individual pulses \citep{CordesMcLaughlin2003} and periodic sources \citep{Morello2020}. Search codes thus generally work under the assumption of a pure Gaussian noise background, which is, however, never realised in practice. The calculation of the detection statistic is dependent on estimating the mean and standard deviation of the background noise from the data; this process is perturbed by the presence of Radio Frequency Interference (RFI), which can act to reduce the estimated signal-to-noise ratio of a genuine source below the detection threshold. It is necessary to subtract away or mask undesirable signals from the data before they are searched, in order to remain acceptably close to the ideal data assumptions and approach the theoretical sensitivity of the instrument.

RFI mitigation can be performed at all stages of the signal chain, with complementary benefits \citep[see e.g.][for an overview]{Baan2010}. One can make a broad distinction between pre-detection techniques, which act on the voltage data stream while it is still available at its highest time resolution, and post-detection methods that operate on the channelised, two-dimensional time-frequency data just before they are searched for radio transients and pulsars. The former are better suited to remove short bursts of RFI with minimal data loss, however they are required to handle large volumes of data in real time and thus tend to be limited to simple thresholding and replacement schemes. A notable exception is that interferometric arrays may implement an additional technique called spatial filtering \citep[e.g.][]{Leshem2000}, where a null in the beam pattern can be placed in the estimated direction of a strong interference source.

On the other hand, the detection stage involves integrating the baseband data in time, making weaker sources of RFI accessible for removal and reducing the data rate, which in turn enables more sophisticated processing. Post-detection methods usually attempt to blank sections of data across the time and/or frequency dimension, based on some specific RFI signatures. Two simple but highly popular methods are the application of a fixed channel mask which permanently discards the most RFI-occupied portions of the band, and the so-called zero-DM filter of \citet{Eatough2009} that aims to subtract away broadband signals that show no cold-plasma dispersion expected from propagation through the interstellar medium. An entire class of algorithms involve flagging sections of data based on whether some statistical property exits the range expected for clean data: for example, a rolling sum over multiple time scales \citep[the \textsc{sumthreshold} algorithm of][]{Offringa2010}, or a higher-order statistical moment \citep{Nita2010}. Convolutional neural networks have also been trained to perform intelligent clipping of time-frequency data \citep[e.g.][]{Akeret2017}. Fourier transforming the data and finding Fourier bins with excessive power is effective against weaker periodic RFI and well-suited to periodicity searches \citep{Fridman2001, Maan2020}.

While there is a trade-off between RFI removal effectiveness and execution speed, the pressure to find methods that perform well on both fronts is increasing, and not all the aforementioned methods are fast enough to be used in a streaming environment. Indeed, on the latest and upcoming generations of radio telescopes, an RFI mitigation algorithm should aim to be:
\begin{itemize}
    \item Fast enough to be used on massively multi-beam systems, where searching large data streams in real-time has now become a fundamental requirement \citep[e.g.][]{CHIMEOverview, SPIEMeerTRAP}
    \item Able to increase the detection probability of genuine sources \textit{and} to reduce the spurious candidate rate at the same time
    \item Transferable between telescopes with limited adjustments to the algorithm's parameters, which would make it more useful to the community as a whole and avoid duplication of work.
\end{itemize}
The challenge lies in developing methods that fulfil all three criteria. 

Static channel masks may be a blunt tool, but are extremely easy to implement and optimal against portions of the band permanently ridden with RFI. The shape and portion of the band occupied by RFI varies as a function of pointing position and of time however; a common situation is that few channels are permanently unusable, while most of them can be occasionally affected with narrow-band RFI; this is arguably the case on MeerKAT for example \citep[][see also \S \ref{subsec:bandpass_tests_meertrap}]{Sihlangu2020}. The ability to automatically and accurately adapt a channel mask to the current data on a time scale of a few seconds is quite desirable; here we propose an algorithm called IQRM that performs this task much faster than real time; IQRM stands for Inter-Quartile range RFI Mitigation, for reasons that will become clear below.

The outline of the paper is as follows. In \S\ref{sec:method}, we explain the rationale behind the design of IQRM, before describing the algorithm in detail. In \S\ref{sec:bandpass_tests}, we test the ability of IQRM to derive a time-variable channel mask from spectral statistics of data obtained with the MeerKAT and Lovell radio telescopes; its execution speed is also measured. We then evaluate the impact of IQRM in a blind search for single pulses in \S\ref{sec:sps_tests}; using over two hours of known pulsar observations taken with the Lovell telescope, we compare the number of genuine astrophysical pulses and false positives reported, with and without IQRM applied. We conclude in \S\ref{sec:discussion} after discussing the performance, limitations and potential extensions to the method.

\section{Method}
\label{sec:method}

\subsection{General principle}
\label{subsec:general_principle}

Search-mode data is a sequence of dynamic spectra recorded with integration times of typically a few tens to hundreds of microseconds. Given a short-duration block of such time-frequency data, the goal of IQRM is to identify all frequency channels affected by narrow-band RFI. The initial step is to reduce the entire block to a single summary statistic per frequency channel (i.e. a spectral moment), and then to determine for which channels the statistic value is unlikely to be associated with clean data.
Nita et al. have previously proposed a method where the data are flagged based on their short-term spectral kurtosis \citep[][]{Nita2007, Nita2010}, for which the acceptable range of values can be determined from first principles. IQRM could be described as a generalisation attempt of the aforementioned work where:
\begin{itemize}
    \item Any spectral statistic can be used as the basis to decide which channels are contaminated by RFI, the only requirement being that higher values must indicate a higher probability of contamination.
    \item The rejection threshold value for the statistic is inferred from the data and assumed to vary slowly with observing frequency; in other words, an arbitrary frequency-dependent background trend or ``baseline" is expected to be present in the sequence of spectral statistic values.
\end{itemize}
Anything that alters the effective system temperature unequally across the band has the potential to generate such a baseline, for example imperfect bandpass filters, variations in receiver response, the diffuse Galactic radio emission \citep[e.g.][]{Haslam1981}, the presence of a bright radio source with a steep spectral index in the telescope field of view, or broadband RFI. The fundamental idea of IQRM is to subtract away the baseline problem by taking the so-called lagged differences of the spectral moment values, and then finding high outliers in these difference values, which is a simpler task.

\subsection{The IQRM algorithm}
\label{subsec:algorithm_statement}

\begin{figure}
    \centering
    \includegraphics[width=1.00\columnwidth]{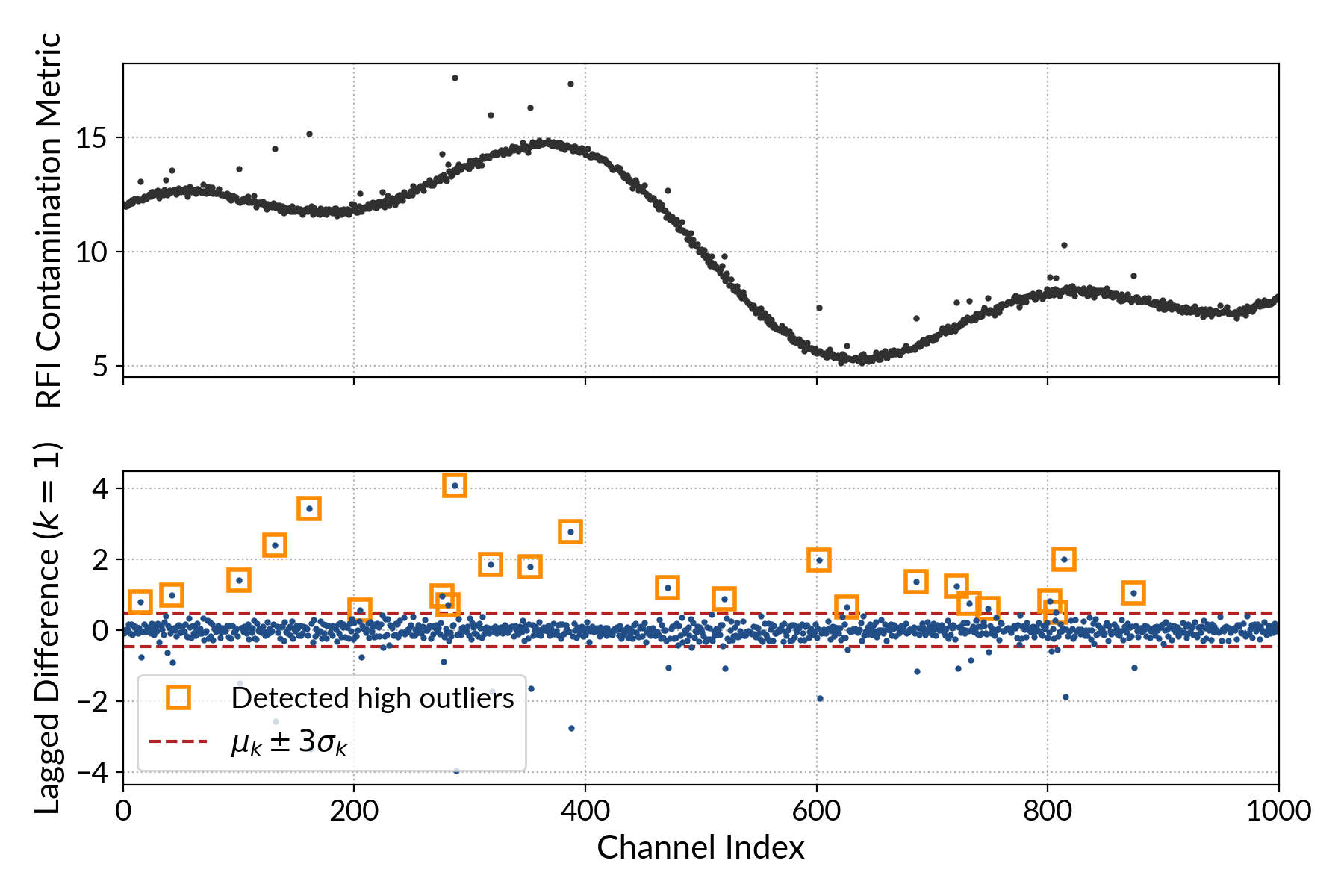}
    \caption{The fundamental idea of IQRM: the lagged difference operation eliminates a slowly varying trend in sequential data and facilitates the procedural identification of outliers. Top panel: artificially generated input data. These are the sum of two sine functions and gaussian noise; an additional 40 points selected at random were then incremented by an exponentially distributed random variate, to simulate high outliers. Bottom panel: lagged difference of the input data (Eq. \ref{eq:lagged_difference}) with a lag of $k=1$ bin; the red dashed lines represent $\mu_k \pm 3 \times \sigma_k$, i.e. the acceptable range of values empirically inferred using the robust mean and scale estimators of Eqs. \ref{eq:robust_mean} and \ref{eq:robust_stddev}. Points highlighted in orange are those deemed abnormally larger than their previous neighbour according to Eq. \ref{eq:outlier_definition}. IQRM repeats this process for multiple values of the lag $k$, in order to overcome cases where consecutive points are outliers with similar values.}
    \label{fig:lagged_diff_trend_elimination}
\end{figure}

Formally, the IQRM algorithm takes as its input an ordered sequence of real-valued numbers $(x_i)_{i=0}^{n-1}$ and identifies values significantly larger than their close neighbours in the sequence. $i$ represents a channel index and $x_i$ the value of the chosen spectral moment (the measure of RFI contamination) in that channel. IQRM has two adjustable parameters:
\begin{itemize}
    \item The radius $r$: the distance (in number of sequence elements) to the furthest neighbour being considered when evaluating the outlier status of a given data point in the sequence.
    \item The threshold $t$: a significance level in number of Gaussian sigmas, which controls by how much a data point must exceed one of its neighbours to be categorised as an outlier. 
\end{itemize}
The first step of the algorithm is to calculate a set of lagged differences of the input sequence $(x_i)$. Here we define the lagged difference of $(x_i)$ for a given lag $k \in \mathbb{Z}$ the sequence $(\Delta_i^k)_{i=0}^{n-1}$ such that

\begin{equation}
\label{eq:lagged_difference}
    \Delta_i^k = x_{i} - x_{i-k},
\end{equation}
where the boundary conditions are handled by setting $x_j = x_0$ for $j < 0$ and $x_j = x_{n-1}$ for $j \geq n$ in the expression above. The usefulness of the lagged difference operation when a trend is present is illustrated in Fig. \ref{fig:lagged_diff_trend_elimination}. It is necessary to calculate lagged differences using multiple trial values for $k$, so that a set of consecutive outliers with very similar values does not escape detection. Trial lag values are selected within the range $[-r, +r]$, excluding $0$. To save computation time, we do not try every possible integer value in that range, but instead arrange the trial values in a geometric progression using the following recurrence relation:

\begin{equation}
\begin{split}
    k_0 &= 1 \\
    k_{m+1} &= \mathrm{max}(\lfloor 1.5 \times k_m \rfloor, k_m + 1),
\end{split}
\end{equation}
where $\lfloor \rfloor$ denotes the floor function, and noting that all $-k_m$ are also added to the sequence of trial values. For example, the sequence corresponding to $r=10$ is $(-9, -6, -4, -3, -2, -1, 1, 2, 3, 4, 6, 9)$. The geometric progression factor of 1.5 has been chosen as a reasonable compromise between masking accuracy and execution speed, but it has not been subjected to a rigorous optimisation process on test data.

\begin{figure*}
    \centering
    \includegraphics[width=1.00\textwidth]{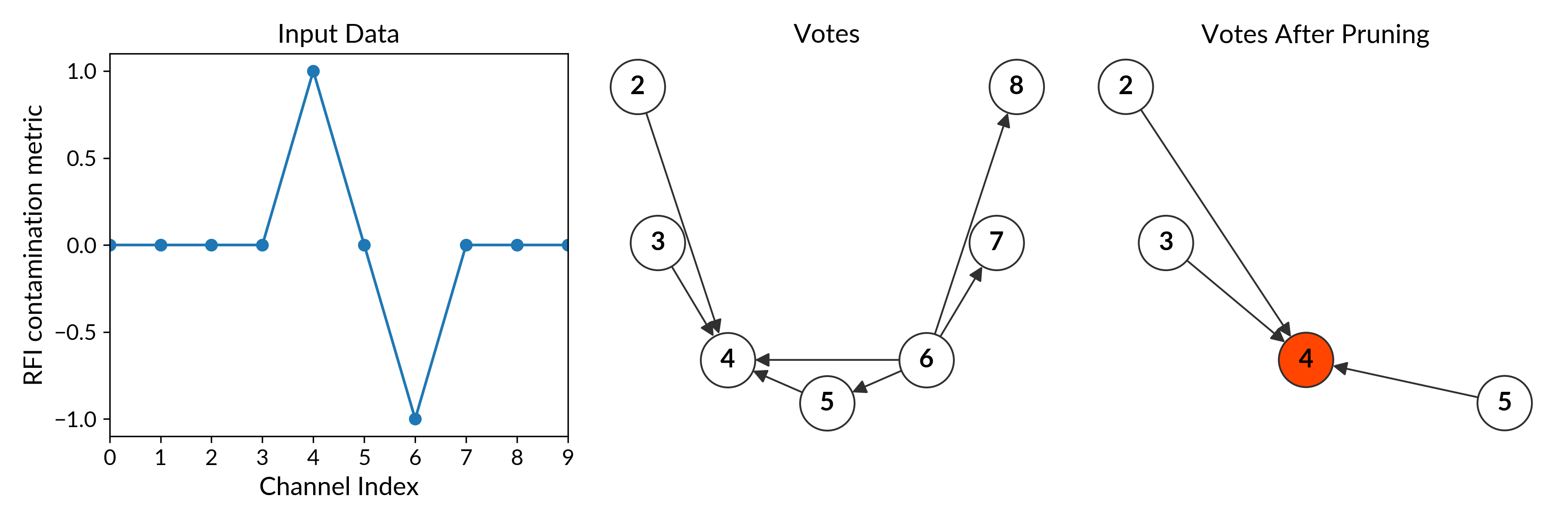}
    \caption{Illustration of the IQRM ``voting" system on a simple test case. Left panel: input data, deliberately generated so that all non-zero lagged differences are considered significant by the algorithm (through Eq. \ref{eq:outlier_definition}). IQRM was then run with a threshold of 3.0 and a radius of 2 channels, which corresponds to trial lag values of \{-2, -1, +1, +2\}. Middle panel: summary of votes cast by channel indices on their neighbours, represented as a directed graph. A vote $i \xrightarrow[]{} j$ is considered valid if and only if $i$ has cast \textit{strictly} less votes in total than $j$ has received; this ensures that neighbours of \textit{low} outliers are not abusively flagged. Here every vote cast by channel 6 is invalid. Right panel: final vote graph after pruning invalid votes; only channel 4 gets marked as a high outlier, reasonably so.}
    \label{fig:iqrm_voting}
\end{figure*}

\begin{figure*}
    \centering
    \includegraphics[width=1.00\textwidth]{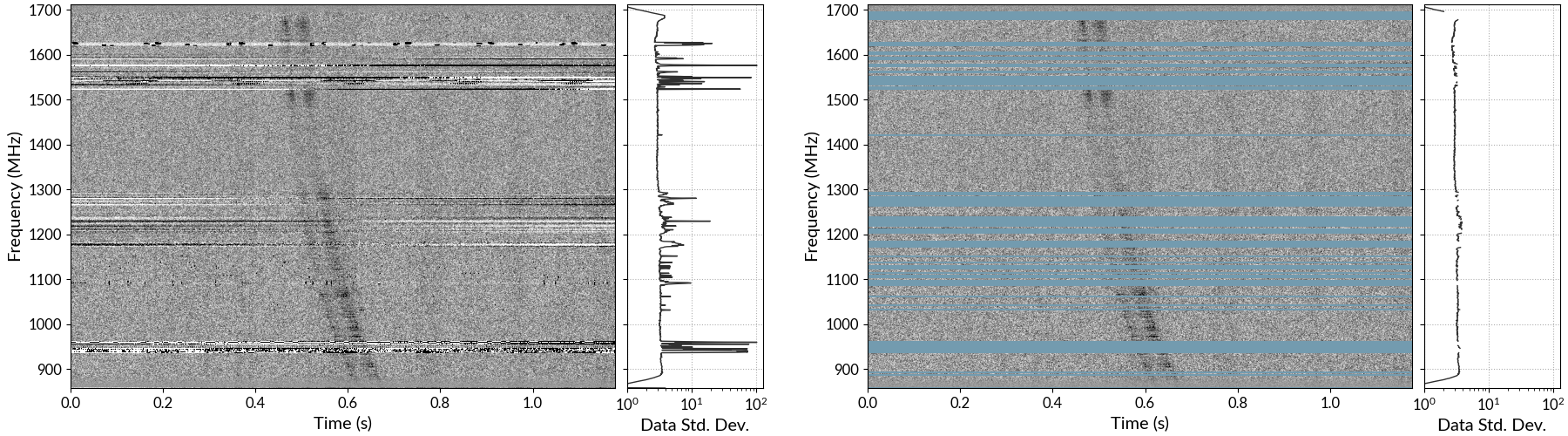}
    \caption{A single pulse from RRAT J1226$-$3223 recorded with MeerKAT at L-Band during the commissioning phase of the MeerTRAP transient search backend (see \S \ref{subsec:bandpass_tests_meertrap} for a system description) in July 2019. Left: original time-frequency data block; here, the spectral standard deviation of the block was provided as the input to IQRM, where the radius parameter was set to $r=100$ and the threshold to $t = 3.0$. Right: same data where the frequency channels flagged by IQRM are shown in blue; the colour scale for valid data and the axis ranges are otherwise unchanged. 17\% of the channels were masked in this example.}
    \label{fig:J1226_before_after}
\end{figure*}

Then, for each trial lag value $k$ separately, \textit{high} outliers in the sequence $(\Delta_i^k)_{i=0}^{n-1}$ are then identified using a criterion similar to Tukey's rule for outliers, also known as Tukey's fences \citep{TukeyFences}. Here we assume that the $\Delta_i^k$ values (for fixed $k$) contain:
\begin{itemize}
    \item A majority of inliers that follow some underlying distribution with mean $\mu_k$ and standard deviation $\sigma_k$.
    \item A minority of outliers that span a range of values much greater than $\sigma_k$
\end{itemize}
$\mu_k$ and $\sigma_k$ need to be measured using robust estimators, i.e. that are not easily biased by outliers. We use the following:

\begin{equation}
\label{eq:robust_mean}
    \mu_k = Q_k(0.5)
\end{equation}

\begin{equation}
\label{eq:robust_stddev}
    \sigma_k = \frac{Q_k(0.75) - Q_k(0.25)}{\Phi^{-1}(0.75) - \Phi^{-1}(0.25)} \approx \frac{\mathrm{IQR}}{1.349}
\end{equation}
where $Q_k(z)$ denotes the $z$-th empirical quantile of the $\Delta_i^k$, and $\Phi^{-1}$ is the inverse cumulative distribution function of the normal distribution. In other words, the mean of the inliers is taken to be the median of the $\Delta_i^k$; the standard deviation of the inliers is taken to be proportional to the empirical inter-quartile range $\mathrm{IQR} = Q_k(0.75) - Q_k(0.25)$, where we used the fact that the inter-quartile range of the normal distribution is $\mathrm{IQR}_\mathrm{norm} = \Phi^{-1}(0.75) - \Phi^{-1}(0.25) \approx 1.349$. For every pair ($i$, $k$), we then test the condition

\begin{equation}
\label{eq:outlier_definition}
    (\Delta_i^k - \mu_k) > t \times \sigma_k,
\end{equation}
which if true denotes that $x_i$ is abnormally larger than $x_{i-k}$, but on its own this does not enable discrimination between cases where $x_i$ is a \textit{high} outlier that should be flagged, and where $x_{i-k}$ is a \textit{low} outlier that should be ignored. Flagging any $x_i$ for which Eq. \ref{eq:outlier_definition} is true for at least one trial lag value $k$ would create a pathological case where the algorithm would mask the entire neighbourhood of every \textit{low} outlier against any common sense. Another processing step, illustrated on a simple example in Fig. \ref{fig:iqrm_voting}, is required to avoid this situation: when Eq. \ref{eq:outlier_definition} is true, we state thereafter that $x_{i-k}$ casts a ``vote" against $x_i$, which we note $(i-k) \xrightarrow[]{} i$. Once collected, the full set of votes can be represented as the edges of a directed graph, where the nodes are array indices. A vote $i \xrightarrow[]{} j$ is considered valid if and only if $i$ has cast \textit{strictly} less votes in total than $j$ has received. Any data point that receives at least one \textit{valid} vote is finally marked as an outlier, and the algorithm returns a binary mask with the same size as the input data sequence. The overall effect of IQRM on real data is shown on an example pulse from a known pulsar recorded with MeerKAT at L-Band (Fig. \ref{fig:J1226_before_after}).


\subsection{Practical implementation and usage}

A minimal open-source implementation of IQRM in the \textsc{Python} language has been made available\footnote{\url{https://github.com/v-morello/iqrm}}. It is minimal in the sense that it provides only the outlier flagging capability specified in \S \ref{subsec:algorithm_statement}. Furthermore, identifying sections of data affected by RFI is not sufficient; this information must then be provided to a search code. The difficulty that arises here comes from the fact that IQRM is meant to generate a channel mask that varies in time: read consecutive data blocks of a few seconds in length, calculate their spectral standard deviation (or other contamination statistic of choice), and obtain a mask adapted to each block by applying IQRM. However, with the notable exception of \textsc{presto} \citep{Ransom2002}, the dedispersion stages of most widely-used search pipelines are not designed to deal with time-variable masks, and instead only accept a fixed list of channels to ignore; such is the case of \textsc{sigproc} \citep{SIGPROC}, \textsc{heimdall}\footnote{\url{https://sourceforge.net/projects/heimdall-astro/}}, \textsc{peasoup}\footnote{\url{https://github.com/ewanbarr/peasoup}}, and \textsc{astroaccelerate} \citep{AASPS}. In many use cases it is thus necessary to replace the bad sections of the original data with adequate substitution values, before sending the edited data for processing. For that purpose, we provide a \textsc{C++} implementation, \textsc{iqrm apollo}\footnote{\url{https://gitlab.com/kmrajwade/iqrm_apollo}}, that performs the full sequence of operations: reading, calculation of the spectral moment, flagging and replacing. It currently processes \textsc{sigproc} filterbank files and saves cleaned copies, noting that interfaces to other formats can be added. There are three possible replacement policies for flagged data:

\begin{enumerate}
    \item Replace by a constant value of the user's choice.
    \item Replace by the mean of the medians of the non-flagged channels within the block.
    \item Replace by artificial Gaussian noise, with mean and variance equal to the global median and global variance of the non-flagged channels within the block.
\end{enumerate}
Mean-of-medians is the default and recommended option, noting that the code is modular and other data replacement strategies can be implemented.

\section{Tests on spectral statistic samples}
\label{sec:bandpass_tests}

In this section we test the efficacy of the IQRM masking process described in the previous section on spectral statistic samples obtained from real-world observations. 

\subsection{MeerTRAP}
\label{subsec:bandpass_tests_meertrap}

\begin{figure*}
    \centering
    \includegraphics[width=1.00\textwidth]{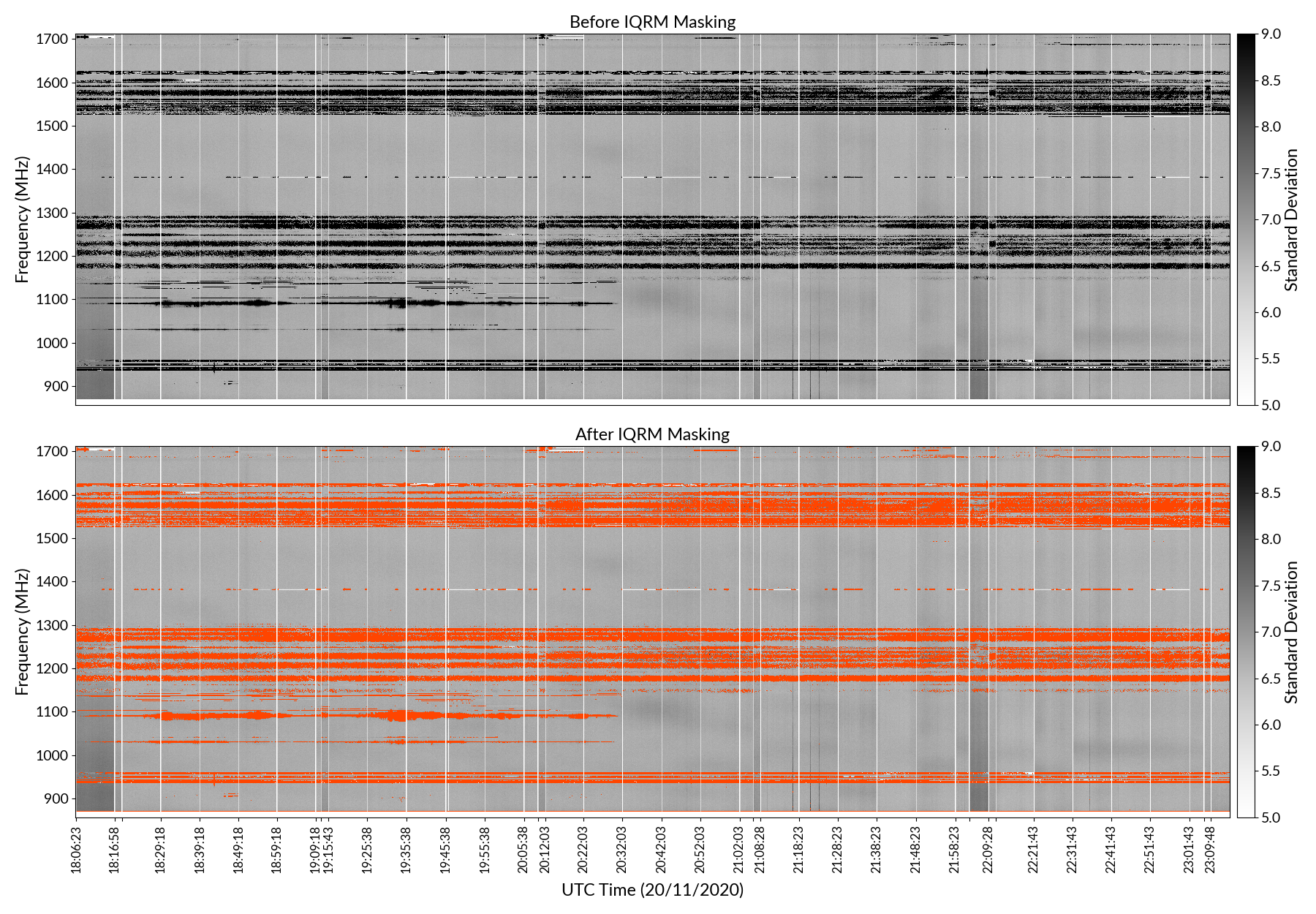}
    \caption{Test of IQRM on 5 hours of MeerTRAP L-Band, 1024-channel data with a 306.24 $\upmu$s sampling interval. Top panel: spectral standard deviation of the data stream from a single tied-array beam as a function of time, measured on consecutive 6-second data blocks. The beamformer aims to produce 8-bit sampled data scaled to a mean of 127 and a standard deviation of 7.0, which in practice happens only in channels dominated by background noise; higher values are a reliable indicator of RFI (see \S \ref{subsec:bandpass_tests_meertrap} for further details). The colour scale has been clipped for readability; some channels can reach a standard deviation $> 100$. Vertical lines correspond to short processing interruptions caused by target changes, and adjustments of the beam tiling pattern on sky that occur every 10 minutes.  Bottom panel: same plot after running IQRM on every spectral standard deviation sample (i.e. every column in the top panel plot), using a radius $r=100$ and a threshold $t=3.0$. Channels flagged by IQRM are shown in orange.}
    \label{fig:meertrap_before_after}
\end{figure*}

MeerKAT is an array of 64 antennas in the Karoo region of South Africa \citep{Jonas2016}, each 13.5 metres in diameter and with the capacity to host four different receivers; both an L-Band (856$-$1712 MHz) and an UHF (544$-$1088 MHz) receiver are currently operational. The raw voltage signals are digitized at the antennas then streamed to the central correlator and beamformer (CBF), which handles channelisation followed by correlation and/or beamforming. The CBF is designed to cover the needs of most telescope users while remaining cost-effective and does not feature a massively multi-beam mode for radio transient searches, which greatly benefit from a wide field of view. However, MeerKAT can host additional backends called USEs (for user-supplied equipment), that may receive a copy of the antenna signals via the CBF and implement additional functionality. MeerTRAP is the commensal, real-time radio transient and pulsar search processor for the MeerKAT radio telescope. It is the association of two sets of user-supplied equipment: 
\begin{itemize}
    \item FBFUSE: a GPU-based beamforming cluster capable of tiling the primary field of view of the telescope with over a thousand tied-array beams~\citep{Barr2018}. 
    \item TUSE: a 66-node search cluster that ingests the beamformed data streams from FBFUSE and searches them for radio transients in real time. The raw data are immediately discarded after processing due to their large volume, and only a limited number of data products are kept. 
\end{itemize}
A more detailed system description can be found in \citet{SPIEMeerTRAP}. The configuration of both components is flexible, but in most observations they are set so that TUSE searches 768 beams tiled in an hexagonal pattern centered on boresight position, and where the data are digitised to 8-bit precision, contain 1024 frequency channels and have a sampling interval of either 306 $\upmu$s (at L-Band) or 480 $\upmu$s (at UHF). In parallel to the search pipeline, each TUSE node runs a so-called bandpass monitor that measures and records spectral statistics of the data for every beam every 6 seconds, which has provided the test dataset for this section. A very important detail to mention here is how the mean and scale of the beamformed data are set on FBFUSE before they are 8-bit digitised. The process is as follows:

\begin{enumerate}
    \item A command to initiate a data rescaling is received, which happens when moving to a new source or at the user's request.
    \item FBFUSE waits until its first input ring buffer block has been filled with data from the CBF. The block has three dimensions: antenna index, observing frequency and time. It contains approximately 0.3 seconds worth of data.
    \item The complex gain corrections are applied to the block.
    \item The mean and standard deviation of the block are calculated for each channel across \textit{all} antennas taken together.
    \item Using these estimates, a scale and an offset term are chosen such that the beamformed data has an expected mean of 127 and standard deviation of 7.0. The scale and offset terms are kept constant and applied to all subsequent blocks until the next rescaling command is received.
\end{enumerate}
The calculation of the spectral scale and offsets assumes that the input data is idealised Gaussian noise: that is, the addition of N signals with unit scale from N independent antennas is expected to result in a beamformed output with a scale of $\sqrt{N}$. However, this assumption breaks down in any channel containing highly directional RFI: in these, the signals from distinct antennas retain a certain level of coherence, and the output scale is thus significantly larger than $\sqrt{N}$. Because of this specific data scaling procedure, the spectral standard deviation of the beamformed data received by TUSE has been found to be an excellent indicator of RFI contamination, and can be passed directly to IQRM. The efficacy of IQRM on MeerKAT is shown in Fig. \ref{fig:meertrap_before_after}, on a 5-hour sequence of spectral standard deviation samples acquired at L-Band. 

We note that a few other spectral statistics were tried in conjunction with IQRM on MeerTRAP data, where the effectiveness of the masking was assessed visually on a sample of known pulsar detections by comparing the data before and after masking as in Fig. \ref{fig:J1226_before_after}. These were the absolute value of spectral skewness, the absolute value of spectral excess kurtosis, and the absolute value of the spectral autocorrelation with a 1-sample delay (see \S \ref{subsec:Lovell} and Eq. \ref{eq:acf1} below for a definition). Absolute values have to be taken for all three to respect the requirement for use with IQRM as stated in \S \ref{subsec:general_principle}, namely that a higher value must indicate a higher probability of contamination. On average, spectral standard deviation was found to leave fewer contaminated channels unmasked. Rigorously exploring more statistics on a larger sample of data remains to be done.

\subsection{Lovell 76-m radio telescope}
\label{subsec:Lovell}
The 76m Lovell telescope (LT) is located at the Jodrell Bank Observatory in the UK. It lies approximately 25 km away from Manchester's city centre and 15 km away from its international airport. The RFI environment is thus particularly adverse compared to MeerKAT or any telescope located in a sparsely populated, radio quiet area. The observing system has been previously described in \citet{Rajwade2020}. The effective observing band spans 336 MHz between 1396 and 1732 MHz divided into 672 frequency channels, and the data are digitised to 8-bit precision. In contrast to MeerTRAP however, the spectral standard deviation of the data does \textit{not} correlate well with RFI contamination here; the variance of the data was often found to be smaller in interference-ridden channels. The most plausible explanation is that the low-noise amplifiers of the receiver operate outside of their linear range in the presence of strong RFI, a regime beyond which their output signal power may become a decreasing function of input signal power, contrary to design expectations (M. Mickaliger, priv. comm.). It was thus necessary to use another spectral moment as an input to IQRM. After testing the same set of spectral statistics that were tried on MeerTRAP data, we found that the best alternative was the absolute value of the spectral autocorrelation with a 1-sample delay, |ACF1| hereafter, defined as:
\begin{equation}
\label{eq:acf1}
    \mathrm{|ACF1|} = \frac{\bigl\lvert  \sum_{i=0}^{n-2} (X_{i+1} - \mu)(X_i - \mu) \bigr\rvert}{n \sigma^2},
\end{equation}
where the $X_i$ denote a sequence of $n$ samples within a given frequency channel, and $\mu$ and $\sigma^2$ are the empirical mean and variance of the $X_i$ respectively. |ACF1| takes values in the range [0, 1] and has all the required properties to be used as a proxy for RFI contamination with IQRM:
\begin{itemize}
    \item Higher values indicate a larger departure from ideal uncorrelated Gaussian noise, for which the expected ACF is 0 regardless of time lag.
    \item It has a computational cost only marginally larger than the spectral standard deviation, which makes it practical for real-time processing.
    \item It is independent from the scale of the data, and thus insensitive to changes in spectral standard deviation caused by the occasional level compression issues faced on LT data.
\end{itemize}

\begin{figure*}
    \centering
    \includegraphics[width=1.00\textwidth]{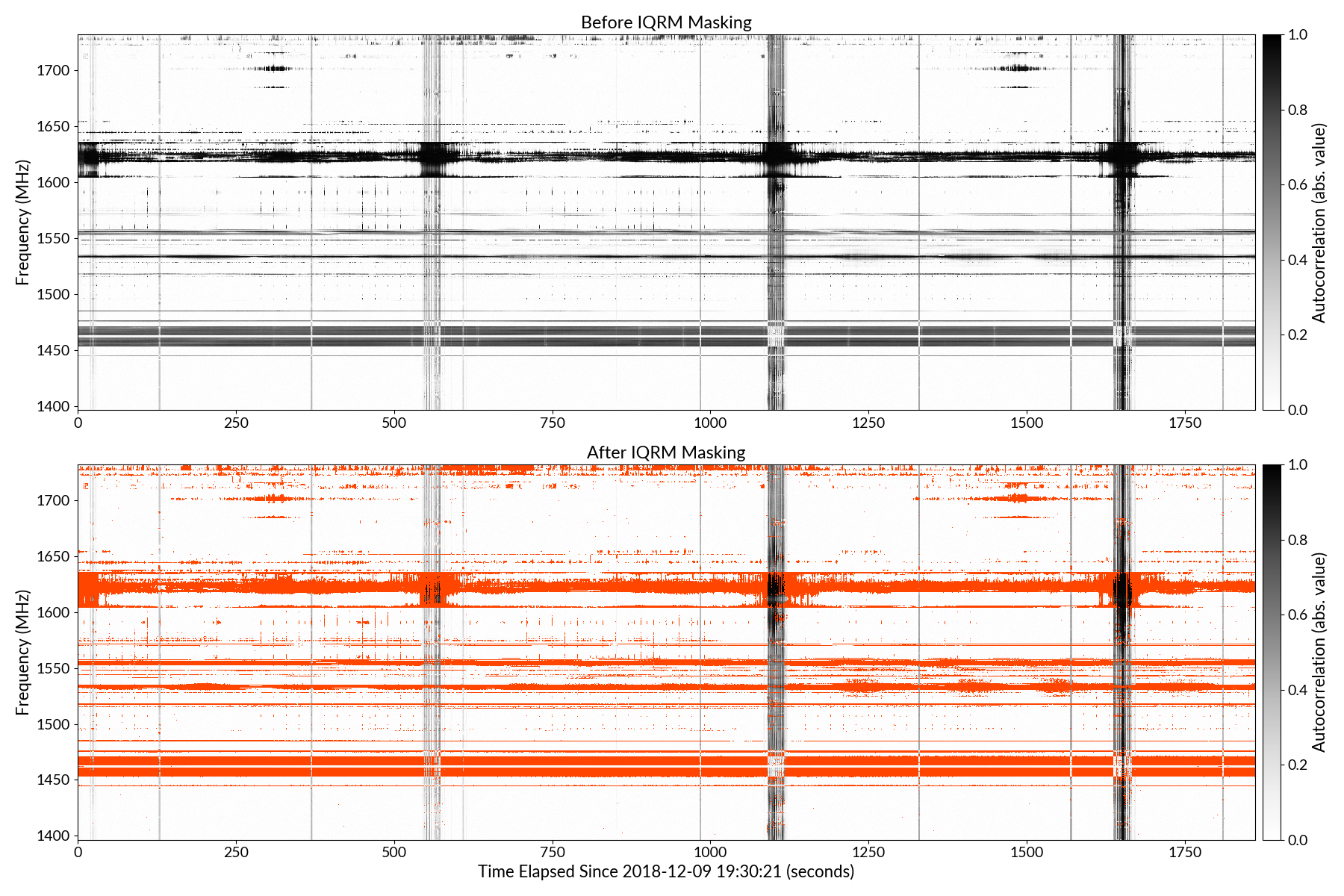}
    \caption{Application of IQRM to a 32-minute Lovell telescope observation of PSR B0611+22, with a 256 $\upmu$s sampling interval and 672 frequency channels. On Lovell data the spectral standard deviation is a much less powerful RFI indicator than on MeerKAT (see \S \ref{subsec:Lovell} for an explanation), and we have thus selected another spectral statistic: the absolute value of the autocorrelation, with a delay of 1 sample. Top panel: spectral autocorrelation measured on consecutive 0.8-second blocks. Bottom panel: same data where IQRM has been applied to every column in the top panel; we used a threshold $t=3.0$, and radius $r=67$, i.e. 10\% of the number of channels. Flagged channels are shown in orange. A pathological case of the algorithm is visible on a few short time intervals here: it fails when a majority of channels are dominated by RFI.}
    \label{fig:jodrell_before_after}
\end{figure*}

A test on a 32-minute sequence of spectral |ACF1| from an observation of PSR B0611$+$22 can be seen in Fig. \ref{fig:jodrell_before_after}, and shows that IQRM can be used with different spectral statistics as an input. There are however short time periods in LT data during which they become essentially worthless across the whole band, as can be seen for example around 1100 seconds after the start of the test observation. This is a case that IQRM cannot properly handle by design, since a fundamental assumption is that the majority of the frequency channels are reasonably clean. In this case, IQRM tends to recommend leaving most or all of the band unmasked, whereas the sensible action would be to discard the data block entirely.

\subsection{Execution speed}

We measured the run times of the \textsc{iqrm apollo} implementation on search-mode data files with different numbers of frequency channels, all containing artificially generated Gaussian noise. These test files were 60~seconds long with a sampling interval of $\tau = 256~\upmu s$, and with a number of channels equal to all powers of two between 256 and 16384 inclusive. We set the IQRM parameters to a radius equal to 10\% of the number of channels and a threshold of $t = 3.0$. The data were processed in blocks of 4096 samples (1.05-second long), and for each block a channel mask was calculated. We ran the code twice on each file, once using spectral standard deviation as the spectral statistic of choice, and once using the spectral autocorrelation with a lag of 1 sample (|ACF1| as described above). In both cases we recorded separately the total time taken by the calculation of the spectral statistic and the total time consumed by the IQRM algorithm \textit{per se}, i.e. the process described in \S \ref{subsec:algorithm_statement}. A single Intel\textregistered~Xeon E5-2630 CPU core was used.

\begin{figure}
    \centering
    \includegraphics[width=1.00\columnwidth]{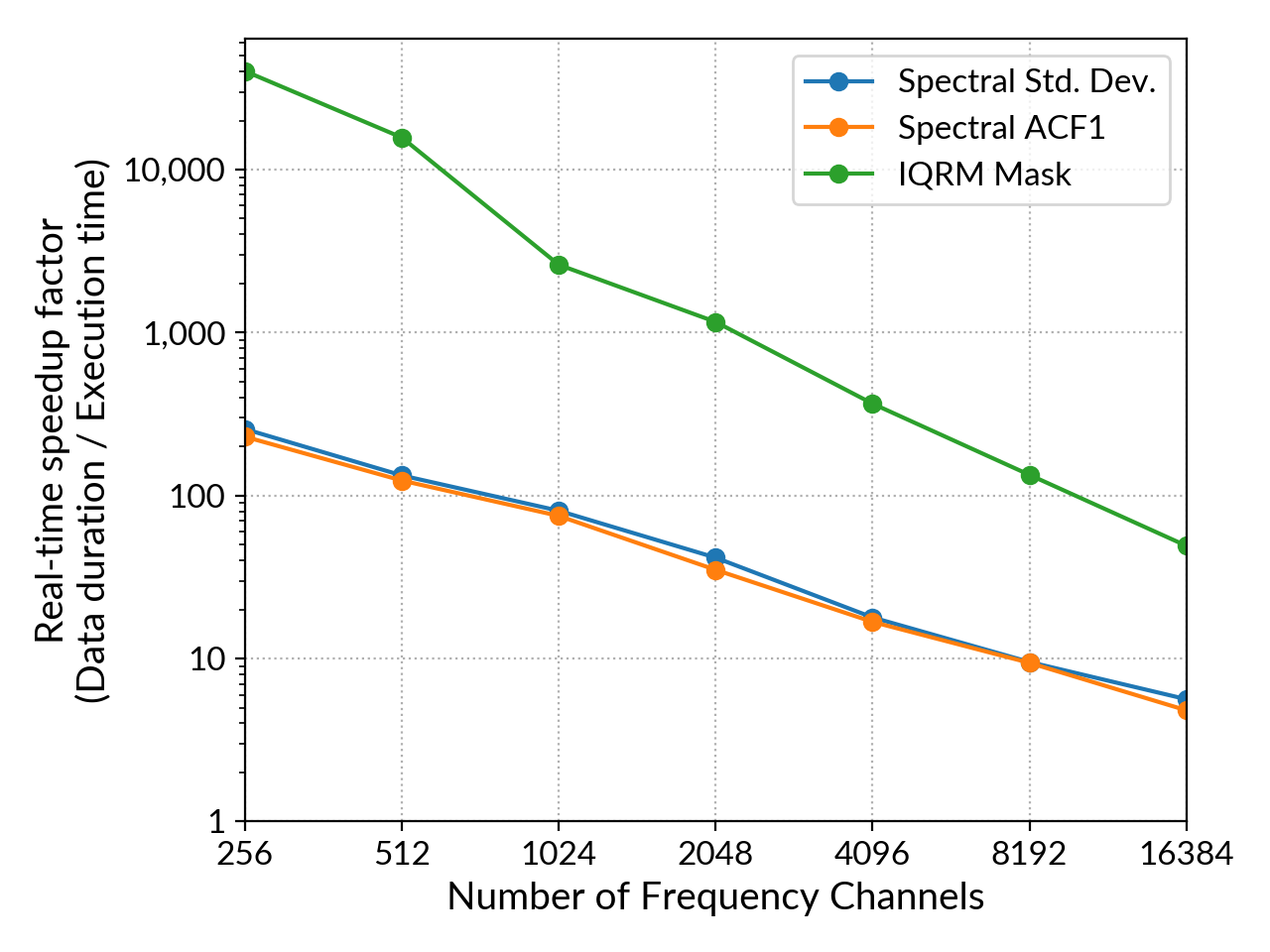}
    \caption{Benchmarks of the \textsc{iqrm apollo} implementation. Using a single CPU core, we ran the code on several test data files containing artificial Gaussian noise, with a sampling time $\tau = 256~\upmu s$ and a number of frequency channels ranging from 128 to 16,384. The data were processed in consecutive blocks of 4096 time samples, approx. 1.05-second long, and a channel mask calculated for each block. We separately measured the average time consumed by the calculation of the spectral statistic (we tested standard deviation and |ACF1| separately), and by the IQRM masking algorithm \textit{per se}. The graph shows the ratio between the effective data length and the execution time of each task, i.e. how much faster than real-time each task would run in a streaming environment.}
    \label{fig:runtime_vs_nchan}
\end{figure}

The benchmarks results can be seen in Fig. \ref{fig:runtime_vs_nchan}; rather than displaying the raw execution times, the ratio between the total data duration and the total run time of each task is shown, which shows how much faster than real time they can be completed in a streaming environment such as MeerTRAP. We note that these results must be linearly extrapolated for a data sampling interval different from $\tau = 256~\upmu s$; indeed the run time scales linearly with the total number of time samples, but not necessarily with the data duration. On both the MeerTRAP and Lovell telescope observing systems described above, IQRM can be run approximately 100 times faster than real-time on a single core of a recent CPU. It should be noted that the calculation of the spectral statistic dominates the total cost, and in principle, this task could be significantly accelerated using many-core architectures compared to the single CPU core used here. A further speedup of one order of magnitude may be achievable assuming that computing resources are plentiful enough.

\section{Tests on single pulse searches with the Lovell telescope}
\label{sec:sps_tests}

Although the experiments done in the previous section are an important visual check of the IQRM algorithm behaviour on its input, the spectral moment passed to the algorithm constitutes only a proxy for interference contamination. Masking all channels for which the spectral moment of choice is deemed high constitutes no guarantee that RFI has been eliminated. The ground truth, that is which exact sections of data are affected by RFI, is never available in practice. The only rigorous way to judge the efficacy of an RFI mitigation method is by measuring how it improves the tangible outcome of the scientific experiment being conducted. In a search for radio transients, one wants to maximize the number of detected pulses from genuine sources while minimizing the number of false positives. In this section, we set out to measure how IQRM improves a standard single pulse search compared to a fixed channel mask, using observations of known pulsars taken with the Lovell telescope.

\subsection{Experimental setup}

\begin{figure}
    \centering
    \includegraphics[width=1.00\columnwidth]{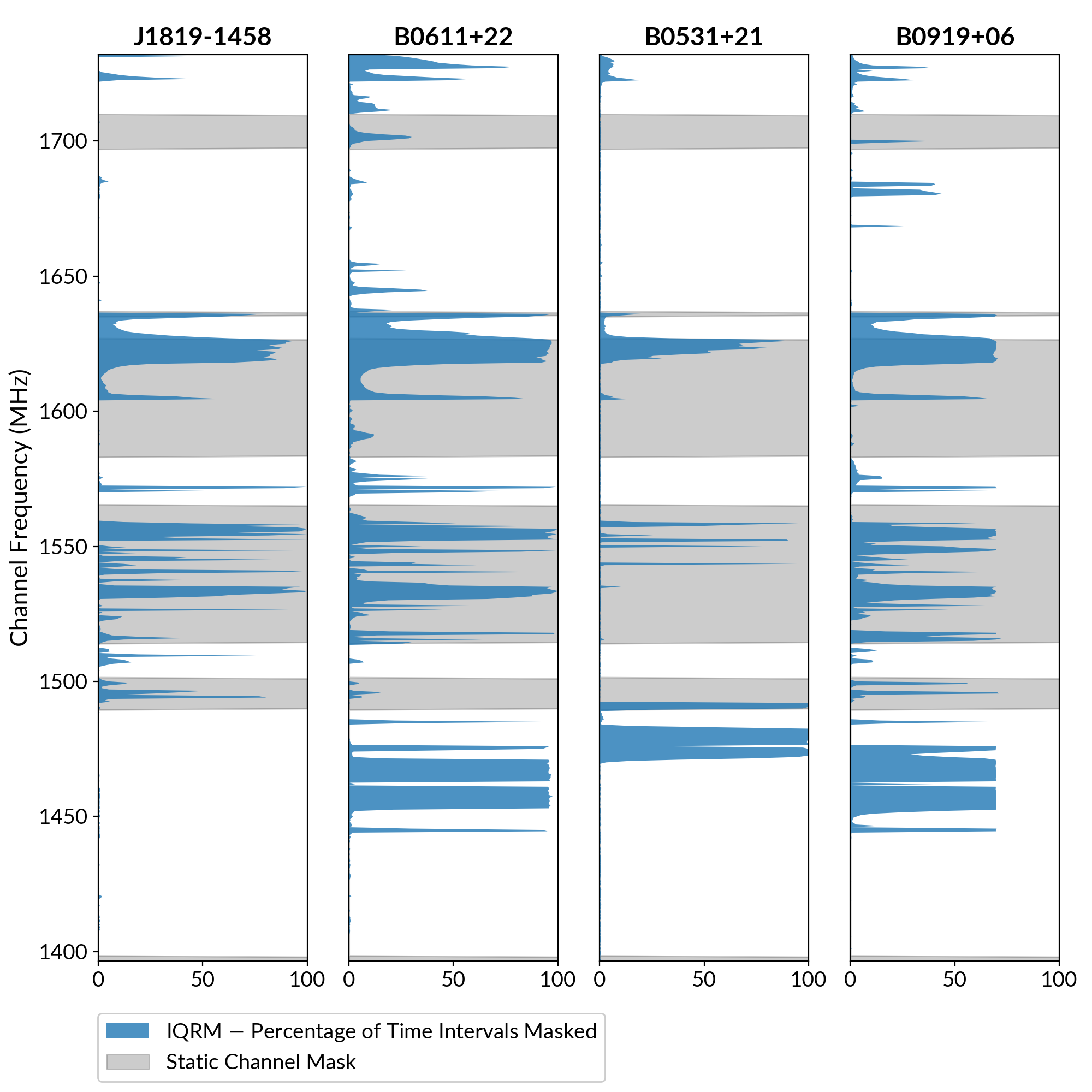}
    \caption{Changes in RFI environment between test pulsar observations taken with the Lovell telescope, underlining the need for adaptive time-dependent RFI masking. The percentage of time intervals masked by IQRM as a function of channel frequency is shown in blue. The static channel mask shown in grey is the one commonly used for Lovell L-band data and was determined prior to, and independently of this work (see text). Note that many RFI-contaminated channels were saturating near the end of the PSR B0919$+$06 observation and were thus left unmasked by IQRM, which explains the apparent clipping of the graph.}
    \label{fig:iqrm_vs_static_mask}
\end{figure}

In order to test the RFI mitigation efficiency of IQRM, we gathered 30-minute long search-mode observations of four known pulsars recorded with the Lovell 76-m radio telescope at 1.4~GHz, with the goal of submitting them to a single pulse search. The observing setup was the same as described above (\S \ref{subsec:Lovell}), with a time sampling interval set to $\tau = 256~\upmu$s. The test sources were PSR~B0611+22, PSR~B0919+06, PSR~J1819-1458 and PSR~B0531+21 (the Crab pulsar), whose parameters are shown in Table \ref{tab:sps_test_pulsars}. These were selected such that we would be able to detect a large sample of individual pulses within the combined two hours of data available, and where a significant fraction of these pulses would be registered with a signal-to-noise ratio near the typical detection threshold of a transient search. The processing pipeline consisted of three stages, following the usual processing model for large-scale radio transient searches:
\begin{itemize}
    \item Search. The data were passed to the GPU-based \textsc{heimdall} search code~\citep{barsdell2012}; the signal-to-noise ratio threshold was set to 6, and the search dispersion measure (DM) range to $[0, 350] \mathrm{pc~cm}^{-3}$. The default internal RFI mitigation options of \textsc{heimdall} were enabled, which consist of narrow-band RFI clipping and a zero-DM filter \citep{Eatough2009}.
    \item Automated candidate classification. The candidates returned by \textsc{heimdall} were passed to \textsc{FETCH} \citep{Agarwal2020}, a radio transient classifier based on a deep convolutional neural network.
    \item Manual candidate vetting. The candidates reported as positive by FETCH were visually inspected; the ones confirmed to be originating from the pulsar were given a final positive label, the others were overruled as negative.
\end{itemize}
On each test data file, we ran two identical instances of the pipeline, which processed copies of the data masked with different methods:
\begin{enumerate}
    \item One instance processed the original data files with only a static channel mask applied. This mask is the ``standard" for the Lovell telescope at 1.4~GHz and was determined independently of the present work, mainly via visual inspection of candidates produced by FRB searches conducted with the same backend \citep[][Mickaliger et al. in prep.]{Rajwade2020}.
    \item The other instance processed copies of the data files that had first been cleaned with the \textsc{iqrm apollo} implementation, where we used |ACF1| as the contamination metric. No static mask was applied.
\end{enumerate}
A comparison of the static channel mask with the average IQRM mask inferred for each test pulsar observation can be found in Fig. \ref{fig:iqrm_vs_static_mask}, where changes of RFI environment between observations are evident.

\begin{table}
    \centering
    \caption{Test pulsars chosen for the single pulse search experiment with the Lovell 76-m telescope at L-Band. DM denotes the dispersion measure and $S_{\mathrm{1400}}$ the average flux density at 1400 MHz. These sources were chosen so that: a) a large sample of detectable single pulses would be recorded in a short amount of observing time (30 minutes for each source), and b) the majority of detectable pulses would be near the typical detection threshold of a search, in a regime where the benefits of improved RFI mitigation are most evident. The data in this table are from the ATNF pulsar catalogue \citep{PSRCAT}.}
    \label{tab:sps_test_pulsars}
    
    \begin{tabular}{lS[table-format=4.1]S[table-format=3.2]S[table-format=2.1]}
        Pulsar           & {Period (ms)} & {DM ($\mathrm{pc~cm}^{-3}$)} & {$S_{\mathrm{1400}}$ (mJy)} \\
        \hline
        PSR B0531$+$21   & 33.4   & 56.77  & 14  \\
        PSR B0611$+$22   & 334.9  & 96.91  & 3.3 \\
        PSR B0919$+$06   & 430.6  & 27.30  & 10  \\
        PSR J1819$-$1458 & 4263.2 & 196.0  & {N/A} \\
        \hline
    \end{tabular}
\end{table}

\subsection{Results}

\begin{figure*}
    \centering
    \includegraphics[width=1.00\textwidth]{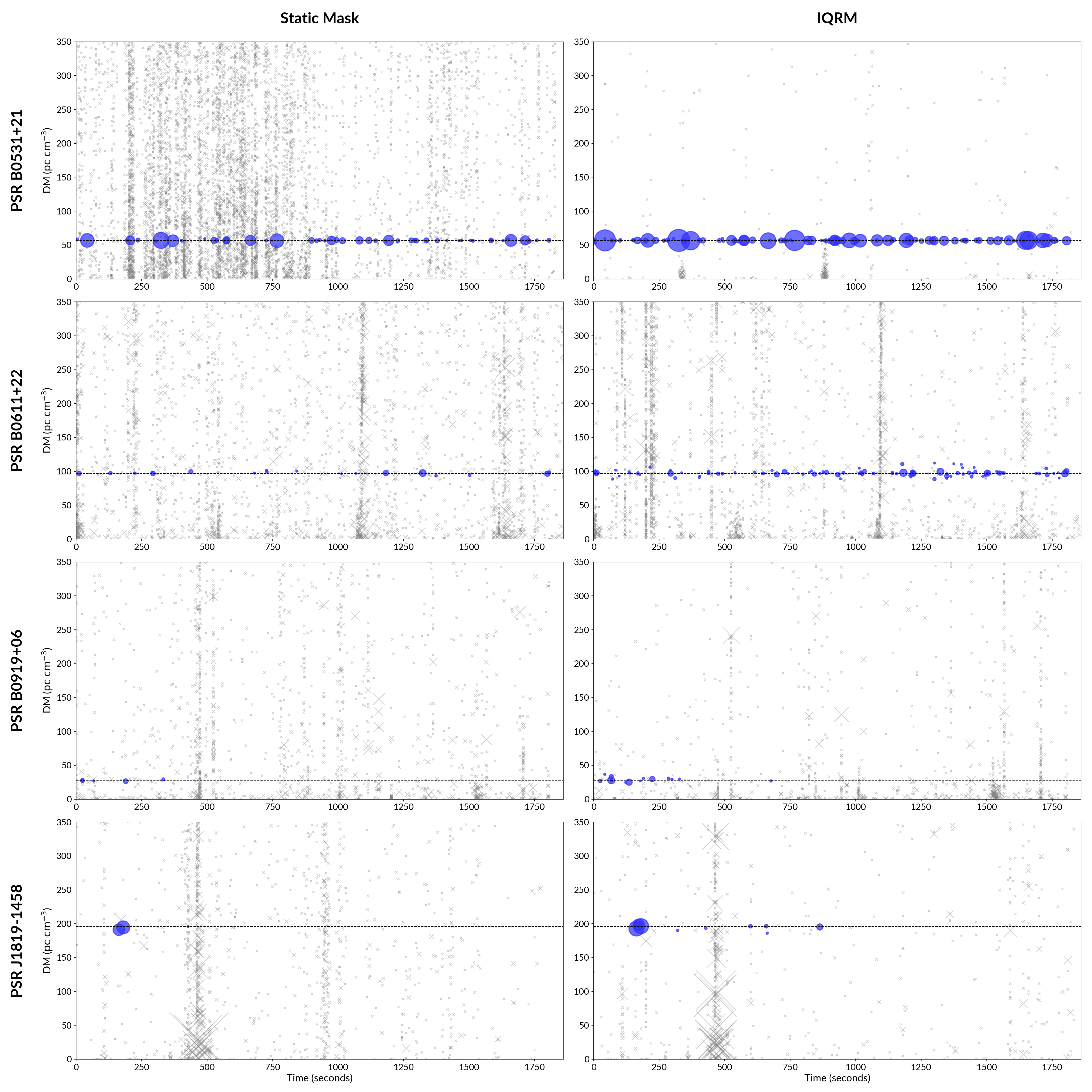}
    \caption{Summary plot of the test single pulse searches. The four known pulsars in Table \ref{tab:sps_test_pulsars} were observed for 30 minutes with the Lovell 76-m telescope at L-Band, and the resulting data were searched with two distinct pipelines: one using a static channel mask, and the other using IQRM instead. Both pipelines were otherwise identical. Confirmed pulse detections are shown as blue circles, with the dispersion measure of the pulsar shown as horizontal dashed lines. Grey crosses correspond to candidates labelled as negatives by the \textsc{fetch} classifier. Time is measured in seconds since the start of each 30-minute observation. All marker sizes are proportional to the signal-to-noise ratio of the event as reported by the \textsc{heimdall} search code. The total number of pulses and of negative candidates for each case is reported in Table \ref{tab:sps_candidate_counts}. Further analysis of which pulses were found by either or both pipelines can be found in Fig. \ref{fig:exclusive_pulse_detections}.}
    \label{fig:sps_summary}
\end{figure*}

\begin{figure*}
    \centering
    \includegraphics[width=0.90\textwidth]{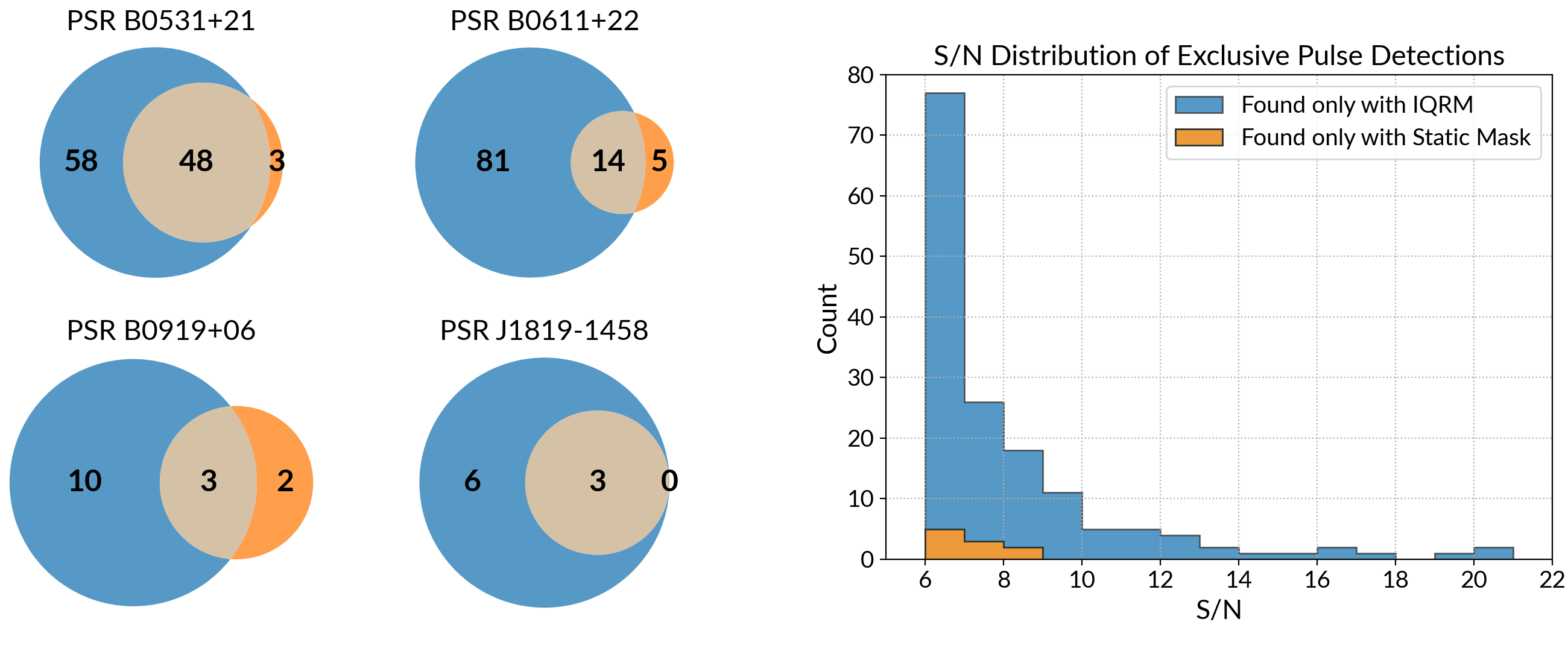}
    \caption{Left panel: Venn diagrams showing the number of confirmed pulses detected by either or both pipelines, for each test source separately. Pulses found exclusively by the IQRM pipeline are shown in blue on the left hand side of each diagram, while those found only by the static mask pipeline are shown in orange on the right hand side. The overlapping area in beige represents pulses detected by both. Right panel: reported signal-to-noise ratio distribution of exclusive pulse detections for all sources combined.}
    \label{fig:exclusive_pulse_detections}
\end{figure*}

The outcome of the search is plotted in detail in Fig. \ref{fig:sps_summary}, where for each pulsar and channel masking method, the candidates reported as positive by FETCH and confirmed to be genuine pulses by visual inspection are shown in a dispersion measure (DM) versus time plane as blue circles. Grey crosses denote the candidates labelled as negatives by FETCH, most of which emerge from RFI that was not entirely masked. Table \ref{tab:sps_candidate_counts} shows the total number of confirmed pulses and of negative candidates produced using either RFI mitigation method. Lastly, the confirmed detections from both pipelines were grouped in time and from there it was determined which pulses were detected by both pipelines, and which were exclusively found by either of them. The outcome of this analysis is shown in Fig. \ref{fig:exclusive_pulse_detections} as a set of Venn diagrams (one for each test source) and a histogram showing the signal-to-noise ratio distribution of the pulses found only by one of the pipelines.

From the overall number of positive and negative candidates, it is apparent that replacing the commonly-used Lovell static channel mask by IQRM is clearly beneficial and that there is no trade-off in doing so, at least in a statistical sense: for every test source, less negative candidates are reported by the search code and more importantly, a significantly larger number of genuine pulses are found. For all sources combined, there is overall a threefold decrease in the negative candidate rate and a nearly threefold increase in the number of astrophysical events detected. However, IQRM does not achieve a perfect result considering that there is a small number of false negatives, i.e. there are 10 pulses in total that were detected with the static mask but not with IQRM. Out of these 10 events, 3 were labelled as negative by the classifier despite having been found with a marginally higher S/N in the IQRM pipeline, while the other 7 were not detected by the search code. After further inspection of these 7 exclusive detections by the static mask pipeline, 3 were found to have been the product of a chance alignment of the pulse with unmasked narrow-band RFI which was removed by IQRM. Considering that masking different channels perturbs both the calculation of the detection statistic and the output value of a machine-learning classifier, it is not reasonable to expect a false negative rate of zero even when switching to a RFI mitigation scheme that is clearly superior on average; however, it is reasonable to expect the false negatives to have signal to noise ratios close to the detection threshold. Here, all the false negatives are arguably faint, the brightest one having been reported with S/N = 8.5. In contrast, a total of 8 pulses with S/N in excess of 15 were missed by the static mask pipeline; all but one are due to mislabeling by the classifier, which was most probably an effect of the residual RFI present in the candidate plots on which it operates.

It is also worth discussing briefly two experimental outcomes that significantly deviate from the average in Table \ref{tab:sps_candidate_counts}. Firstly, the larger than average increase in genuine pulse detections for PSR B0611$+$22 is worth mentioning. We attribute it to the fact that, during this observation, the vast majority of pulses from the source had signal-to-noise ratios (SNRs) concentrated near the threshold of the search code, which was set to 6. In this regime, the consistent increase in SNR provided by the use of IQRM pushed dozens of pulses just above the detection threshold. Secondly, the exceptionally large decrease of 96\% in negative candidates in the PSR B0531$+$21 observation; closer inspection of the data showed that in this case, most candidates produced in the static mask run are caused by the RFI source occupying the 1450$-$1470 MHz frequency range, which manifests itself as a square wave with a period usually equal to two time samples. This interference source was active in other observations, for example in the PSR B0611$+$22 one, which can be seen in Fig. \ref{fig:jodrell_before_after}, where it did not cause a significant portion of the spurious candidates; however, it was unsually bright during the PSR B0531$+$21 observation, and the fact that the incriminated frequency band is \textit{not} part of the static channel mask had a significant impact. In contrast, IQRM nearly always masked this portion of the spectrum.

\begin{table*}
    \centering
    \caption{Summary of candidate numbers produced by the test single pulse searches. Using IQRM instead of the static channel mask resulted both in a reduction in the number of spurious candidates reported, and in a significantly increased number of astrophysical events identified.}
    \label{tab:sps_candidate_counts}
    \begin{tabular}{lllllll}
                     & \multicolumn{3}{c}{\textbf{Confirmed pulse detections}} & \multicolumn{3}{c}{\textbf{Negative candidates}}  \\
    Source           & Static Mask   & IQRM           & Change                 & Static Mask    & IQRM          & Change           \\ \hline
    PSR B0531$+$21   & 51            & 106            & $+108\%$               & 7079           & 312           & $-$96\%          \\
    PSR B0611$+$22   & 19            & 95             & $+400\%$               & 2883           & 2423          & $-$16\%          \\
    PSR B0919$+$06   & 5             & 13             & $+160\%$               & 1442           & 941           & $-$35\%          \\
    PSR J1819$-$1458 & 3             & 9              & $+200\%$               & 1134           & 629           & $-$45\%          \\ \hline
    \textbf{Total}   & \textbf{78}   & \textbf{223}   & $\mathbf{+186\%}$      & \textbf{12538} & \textbf{4305} & $\mathbf{-66\%}$ \\
    \end{tabular}
\end{table*}

\section{Discussion and Conclusion}
\label{sec:discussion}

Implementing effective RFI mitigation is now a necessity on any radio telescope; the task is made even more challenging on recent massively multi-beam systems where real-time processing is required. The widely-used static channel masks cannot effectively cope with the RFI environment variations as a function of pointing position or time of the day for example. We have introduced IQRM, an algorithm that infers a time-dependent, adaptive channel mask directly from consecutive blocks of search-mode data. Our results show that it fulfills all three desirable characteristics for use in real-time searches with modern massively multi-beam systems. Firstly, it runs much faster than real-time (approximately 100 times faster on Lovell telescope data, with a single CPU core) and it should thus be possible to integrate it into any existing data processing pipeline without measurably affecting the speed of the whole search process. Secondly, in a radio transient search, IQRM is able to largely reduce the number of false positives but also enhance the S/N of genuine astrophysical events; on Lovell telescope data, the number of individual pulses detected from a set of four known pulsars increased by an average factor of 3 when replacing the standard, commonly-used static channel mask by IQRM. Lastly, the algorithm is non-parametric and should thus transfer well between different telescopes. It must be noted however that the transfer is not entirely trivial, as one important task is still left to the user: finding a spectral moment or statistic that correlates well with the presence of RFI on the observing system at hand. Our analysis above indicates that the correct choice depends chiefly on the dynamic range of the input data. We found that on MeerKAT, the specific method used to determined the data digitisation levels is such that the standard deviation of the data in a frequency channel is an excellent proxy for RFI contamination, and a natural choice as the IQRM input. On the Lovell telescope however, the dynamic range of the data was found to be unexpectedly \textit{reduced} in the presence of strong RFI bursts, due to a technical limitation of the low-noise amplifiers; this motivated using the spectral autocorrelation of the data as an input to IQRM instead. With that in mind, the IQRM \textsc{C++} implementation provides multiple spectral statistics to choose from. The code is modular and users may add new ones if need be.

Much like all existing techniques, IQRM is not universally effective against all forms of RFI and is primarily used as a bright narrow-band interference masking algorithm. Its main limitation is that the estimator it uses to infer the acceptable range for its input spectral statistic (Eq. \ref{eq:outlier_definition}) implicitly assumes that at least 50\% of the channels are clean at any point in time; the method becomes less effective otherwise and masks less data than expected, a situation that was found to occasionally occur on Lovell telescope data. A possible solution would be to add a short-term memory to the algorithm, i.e. take into account the spectral statistics of past data blocks when determining the acceptable range for said statistic in the most recent block; this could help in making the correct decision when all channels become temporarily contaminated (see Fig. \ref{fig:jodrell_before_after}). Another potential application of IQRM would be to use the outlier detection algorithm on the data integrated in time, i.e. the zero-DM time series. This could make it useful against broadband, non-dispersed RFI pulses, much like the now widely used zero-DM filter devised by \citet{Eatough2009}, but without systematically reducing the S/N of genuine, low DM astrophysical events. The two aforementioned extensions will be considered for a later update of the \textsc{iqrm apollo} implementation.

\section*{Acknowledgements}

VM, KMR and BWS acknowledge funding from the European Research Council (ERC) under the European Union's Horizon 2020 research and innovation programme (grant agreement No. 694745). The authors thank Mitchell Mickaliger for useful technical discussions about the Lovell telescope backend. The authors would  like  to  thank  the  South African Radio Astronomy Observatory (SARAO)  for  the  approval to use MeerKAT data for the analysis presented in this paper. The MeerKAT telescope is operated by SARAO, which is a facility of the National Research Foundation, an agency of the Department of Science and Innovation. Pulsar research at Jodrell Bank and data acquisition with the Lovell Telescope is supported by a consolidated grant from the UK Science and Technology Facilities Council (STFC). We thank the anonymous referee and the editor Tim Pearson for their suggestions that helped significantly improve the quality of this article. We also thank Scott Ransom for pointing out an erroneous statement about \textsc{presto} in the initial version of the paper (\textsc{presto} can in fact take a time-variable channel mask as an input, contrary to what was previously claimed).


\section*{Data Availability}
The data used in the analysis presented here are publicly available in a repository on Zenodo\footnote{\url{https://doi.org/10.5281/zenodo.5256885}}.



\bibliographystyle{mnras}
\bibliography{references}







\bsp	
\label{lastpage}
\end{document}